# Common Software for the ALMA project

G. Chiozzi, B. Gustafsson, B. Jeram, P.Sivera - ESO, Garching bei Muenchen, DE
M. Plesko, M. Sekoranja, G. Tkacik, J. Dovc, M. Kadunc, G. Milcinski, I.Verstovsek, K. Zagar - JSI, Ljubljana, SI

Abstract

The Atacama Large Millimeter Array (ALMA) is a joint project between astronomical organizations in Europe, USA and Japan. ALMA will consist of at least 64 12-meter antennas operating in the millimeter and sub-millimeter wavelength range, with baselines up to 10 km. It will be located at an altitude above 5000m in the Chilean Atacama desert[1].

The ALMA Common Software (ACS) provides a software infrastructure common to all partners and consists of a documented collection of common patterns in control systems and of components, which implement those patterns. The heart of ACS is an object model of controlled devices, called Distributed Objects (DOs), implemented as CORBA network objects. Components such as antenna mount, power supply, etc. are defined by means of DOs. A code generator creates Java Bean components for each DO. Programmers can write Java client applications by connecting those Beans with data-manipulation and visualization Beans using commercial visual development tools or programmatically.

ACS is based on the experience accumulated with similar projects in the astronomical and particle accelerator contexts, reusing and extending concepts and components. Although designed for ALMA, ACS has the potential for being used in other new control systems and other distributed software projects, since it implements proven design patterns using state of the art, stable and reliable technology.

## 1 INTRODUCTION

Since the beginning of the ALMA project we have been aware of the complexity of the project with respect to geographically distributed development and differences in the traditional way of working of the teams involved. The number of applications and developers will increase to very large numbers.

To alleviate these problems, we have decided to introduce a central object oriented framework, that we call the ALMA Common Software (ACS). It is located in between the ALMA application software and other basic commercial or shared software on top of the operating systems. It provides a well-tested platform that embeds standard design patterns and avoids duplication of effort. At the same time it is a natural platform where upgrades can be incorporated and brought to all developers. It also allows, through the use of well-known standard constructs and components, that other team members that are not authors of ACS easily understand the architecture of software modules, making maintenance affordable even on a very large project.

In order to avoid starting from scratch, we have evaluated emerging systems that could provide a good basis for ACS and bring in the project CORBA and other new technology know-how. We have then started a fruitful collaboration between ESO and JSI that, through cross-fertilization of experience and ideas from our previous projects has driven us to the concepts and implementation of ACS. For more details on the considerations that have led to the concepts behind ACS see [2] and [3][4].

## 2 ACS ARCHITECTURE

ACS is based on the object oriented CORBA middleware, which gives the whole infrastructure for the exchange of messages between distributed objects and system wide services [5]. Whenever possible, ACS features are implemented using off-the-shelf components; ACS itself provides in this case the packaging and the glue between these components.
The object paradigm of CORBA is fully used: each

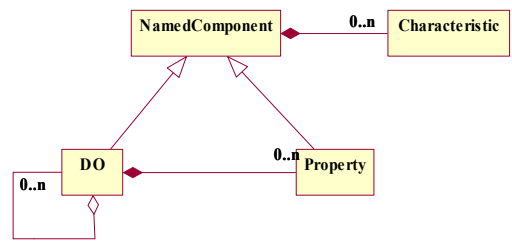

Figure 1: DO – Property – Characteristic class diagram

controlled device type is represented by one specific CORBA interface that subclasses the base distributed object (DO). Each DO is further composed of properties that correspond to what is called controlled points, channels or tags in SCADA systems. Each property is an object too, implementing get/set commands, event-driven monitors and alarms, asynchronous/synchronous communication, describing

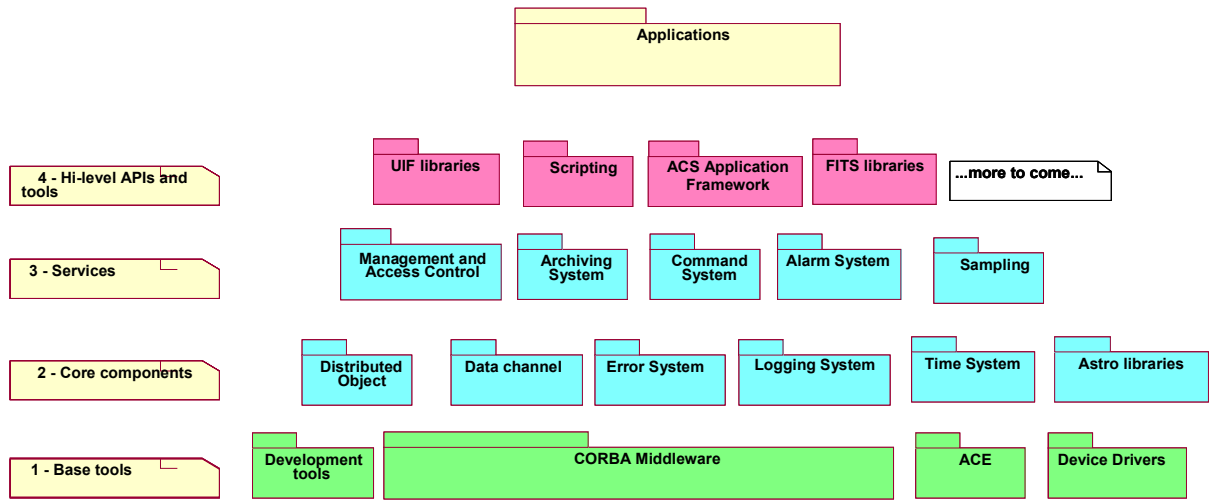

Figure 2: ACS Packages

itself via characteristics such as min/max, units, etc. (figure 1).

The UML Package Diagram in Figure 2 shows the main packages in which ACS has been subdivided. For more details, refer to the ACS Architecture, available on the ACS Web Page[12].

## 2.1 Base Tools

The bottom layer contains base tools that are distributed as part of ACS to provide a uniform development and run time environment on top of the operating system for all higher layers and applications. These are essentially off-the-shelf components and ACS itself just provides packaging, installation and distribution support. This ensures that all installations of ACS (development and run-time) will have the same basic set of tools.

## 2.2 Core components

This second layer ensures standard interface patterns for all distributed objects and provides essential components, necessary for the development of any application. Among these:

- **Distributed Object (BACI)**
  Base interfaces and classes for Distributed Object, Properties and Characteristics (see figure 1) are implemented in this package. This component is called Basic Control Interface (BACI) [6].
- **Data Channel**
  The Data Channel provides a generic mechanism to asynchronously pass information between data publishers and data subscribers, in a many-to-many relation scheme.
- **Time System**
  Time and synchronization services.
- **Error System**
  API for handling and logging run-time errors, tools for defining error conditions; tools for browsing and analyzing run-time errors.
- **Logging System**
  API for logging data, actions and events. Transport of logs from the producer to the central archive. Tools for browsing logs.

## 2.3 Services

The third layer implements higher level services. Among these:

- **Management and access control interface (MACI)**
  Design patterns, protocols and high level meta-services for centralizing access to ACS services and DOs, to manage the full life-cycle of DOs including persistent store, and to supervise the state of the system [7]
- **Archiving System**
  API tools and services for archiving monitoring data and events.

## 2.4 API and High-level tools

The fourth and last layer provides high level APIs and tools. The main goal for these packages is to offer a clear path for the implementation of applications,

with the goal of obtaining implicit conformity to design standards. Among these, we mention:

- **UIF Libraries**
  Development tools and widget libraries for User Interface development. Java user interfaces are based on the ABeans library that wraps CORBA objects into Java Beans, which are then connected with commercial data-manipulation and visualization Beans using visual tools or programmatically[8].
- **ACS Application Framework**
  Implementation of design patterns and to allow the development of standard applications.

## 3 ACS DEVELOPMENT STATUS

The development of ACS is driven by the needs of the teams developing higher level software, and in particular the ALMA Control System.

Our development cycle foresees one major release every year, with an intermediate, bug-fixing, release after six months.

ACS 0.0, released in September 2000 was essentially a concept demonstration prototype. With the support of some components of the VLT Control Software[3][10] it has been used to develop a prototype control system for the 12m Kitt Peak antenna. This was successfully tested in December 2000.

ACS 1.0, released in September 2001, is the first "production release". It is used for the development of TICS, the ALMA Test Interferometer Control System[11] (a first prototype was developed with ACS 0.0 in the first half of 2001). Two test antennas will be installed starting from spring 2002 at the VLA site.

The next major release, ACS 2.0, is foreseen for September 2002 and will be a fully functional system.

ACS will then need to be extended to accommodate the needs of the ALMA Data Flow Subsystems (Archiving, Scheduling, Observation Tools, Pipeline, etc.)

## 4 CONCLUSION

ACS has been developed keeping in mind the needs of a wide range of astronomical and accelerator control projects. It can easily run on many platforms and operating systems and is open source. The complete code is compiled with standard GNU cpp, including the sources of the underlying CORBA implementation, TAO[13], which is also open source. A part of the service client applications are written in Java, using ORBacus[14] for the ORB, which is free for non-commercial purposes. ACS is currently supported on Linux and VxWorks. A Solaris and a MS Windows version are also used internally at ESO and JSI for testing purposes.

We are therefore convinced that many other projects can use ACS. At the same time, we think that a wider user's base can provide us with very valuable feedback.

## 5 ACKNOWLEDGEMENTS

The ACS project is managed by ESO in collaboration with JSI. This work is the result of many hours of discussions, test and development inside our groups and in the various ALMA centers at NRAO, IRAM and Bochum. We thank here all our colleagues for the important contribution to the definition and implementation of ACS.